\documentclass[%
 reprint,
groupedaddress,
showpacs,
 amsmath,amssymb,
 aps,
prb,
]{revtex4-1}

\usepackage{graphicx}
\usepackage{dcolumn}
\usepackage{bm}

\begin{document}

\newcommand{\be}{\begin{equation}}
\newcommand{\ee}[1]{\label{#1}\end{equation}}
\newcommand{\bem}{\begin{eqnarray}}
\newcommand{\eem}[1]{\label{#1}\end{eqnarray}}
\newcommand{\eq}[1]{Eq.~(\ref{#1})}
\newcommand{\Eq}[1]{Equation~(\ref{#1})}
\newcommand{\vp}[2]{[\mathbf{#1} \times \mathbf{#2}]}


\title{Transverse forces on vortices in superfluids in a periodic potential}

\author{E.  B. Sonin}
 \affiliation{Racah Institute of Physics, Hebrew University of
Jerusalem, Givat Ram, Jerusalem 91904, Israel}

\date{\today}

\begin{abstract}
The paper analyzes the transverse forces (the Magnus and the Lorentz forces) on  vortices in superfluids put into periodic potentials at $T=0$. The case of weak potential and the tight-binding limit described  by the Bose--Hubbard model were addressed. 
The analysis was based on the balance of true momentum and quasimomentum.  A special attention was paid to the superfluid close to the superfluid-insulator transition. In this area of the phase diagram  the theory predicts the particle-hole symmetry line  where the Magnus force changes  sign with respect to that expected from the sign of velocity circulation. Our analysis has shown that the magnitude of the Magnus force is a continuous function at 
crossing the particle-hole symmetry line. This challenges the theory connecting the magnitude of the Magnus force with topological Chern numbers and predicting a jump at crossing this line. Disagreement is explained by the role of intrinsic pinning and guided vortex motion ignored in the topological approach.
It is one more evidence that in general topological arguments are not sufficient for derivation of equations of vortex motion. 
   \end{abstract}

\pacs{47.32.C-,03.75.Lm,47.37.+q}
\maketitle


\section{Introduction} \label{Intr}

The transverse force on a vortex in superfluids (neutral and charged)  is debated during many decades and has been a topic of reviews and books \cite{RMP,Kop,VolB,EBS}. In a Galilean invariant continuous superfluid at $T=0$, the balance of forces on a vortex is 
\be
\bm F_M + \bm F_L= \bm F_e,
     \ee{B}
where 
the Magnus force  
\be 
\bm F_M= n m[\bm  v_L  \times \bm  \kappa].
       \ee{Mfg}
is proportional to the vortex velocity $\bm v_L$ and the Lorentz force 
\be 
\bm F_L=- [\bm  j  \times \bm  \kappa],
       \ee{Lf}
 is proportional to the transport superfluid mass current $\bm j =mn \bm v_s$. Here $n$ is the particle density  with mass $m$, $\bm \kappa$ is the vector parallel to the vortex axis with its modulus equal to the circulation quantum $\kappa=h/m$, and the superfluid velocity $\bm v_s ={\hbar \over m} \bm \nabla \theta$ is determined by the phase $\theta$ of the order parameter wave function. The external force $ \bm F_e$ combines all other forces on the vortex, e.g., pinning and mutual friction forces.  The total transverse force $\bm F_M + \bm F_L$ depends only on the relative velocity $\bm v_L-\bm v_s$ as required by the Galilean invariance. If the external force $\bm $ is absent the vortex moves with the superfluid velocity ($\bm v_L=\bm v_s$) as required by Helmholtz's theorem.

The Magnus force leads to the Hall effect if particles  have a charge $q$. The electric field is determined by the vortex velocity: ${\cal \bm E}={1\over c} [\bm B\times \bm v_L]$.  The value of the Hall conductivity 
\be
\sigma_H= {j_q \over {\cal E}}={q n c\over B } {v_{cm}\over  v_L },
   \ee{QHC}
where  $j_q ={q\over m}j$   is the charge current and 
\be 
v_{cm} ={j \over mn} 
    \ee{cmv}
is the center-of-mass velocity, which may differ from the superfluid velocity $v_s$ in general. Thus the ratio $v_{cm}/v_L$ fully determines the Hall conductivity.
If the vortex moves with the  center-of-mass velocity (Helmholtz's theorem) the Hall conductivity has the universal value 
\be
\tilde \sigma_H={qnc \over B}.
   \ee{sHu}

If a superfluid is in a periodic potential the Galilean invariance is broken, and  the value of the Magnus force   was a matter of debates, especially in the limit of strong periodic potential (tight-binding limit) when the Bloch band theory reduces to a lattice model. 
The most known lattice model of the superfluid is the Josephson junction array. In the classical theory of the Josephson junction array usually the particle-hole symmetry is assumed, which  forbids the Magnus force in the model (see Ref.~\onlinecite{EBS} and references therein). However, this symmetry is not exact, and there  was a lot of theoretical works aiming at finding  a finite Magnus force, mostly suggesting some quantum effects.

Intensive investigations of Bose-condensed cold atoms attracted an interest to another lattice model of a superfluid:  the Bose--Hubbard model. The periodic structure of potential wells for bosons, which leads to the Bose--Hubbard model in the tight-binding limit,  is realized for cold-atom BEC in experiments with optical lattices \cite{Ued}. Lindner {\em et al.} \cite{Auer} and Huber and Lindner \cite{Lind} calculated the Magnus force in the Bose--Hubbard model and revealed that close to the superfluid-insulator transition the force changes its sign as happens in Fermi superfluids at changing the sign of the carrier charge. Berg {\em et al.}\cite{Berg} extended this theory to charged Fermi superfluids (superconductors). All of them used topological arguments connecting the Magnus force with Chern numbers, and the theory predicted quantized Hall conductivity multiple of the value given by \eq{sHu}. This led to a conclusion\cite{Auer,Lind} that at the particle-hole symmetry line  the Magnus force changes its sign not continuously but with a jump in its magnitude.

 This analysis was challenged\cite{Son13a,EBS} from the position that derivation of any force must rely first of all on the  momentum balance, which the authors of Refs.~\onlinecite{Auer,Lind} did not consider. The analysis based on this concept has shown that there is no jump in the Magnus force magnitude at the particle-hole symmetry line. However, the momentum balance analysis of Refs.~\onlinecite{Son13a,EBS}   was not free from risky assumptions and did not go beyond the coarse-grained treatment valid only at distances much longer that the period of the lattice. 

The present paper overcomes these shortcomings. The new analysis demonstrates  that the coarse-grained model is not sufficient indeed. Moreover, a credible estimation of the Magnus force (and also Lorentz force) is impossible within the lattice model such as the Bose--Hubbard model. One must analyze a microscopic wave function defined in the whole space but not only on discrete sites of the lattice. According to the present analysis, the very concept of a force on a vortex is ambiguous since ambiguous is definition of the vortex displacement. The vortex is not a rigid object and any its displacement in a periodic potential is accompanied by its deformation. However, this ambiguity is not a serious hurdle since it does not affect the ratio of the center-of-mass velocity $v_{cm}$ and the vortex velocity $v_L$. The latter is determined by the ratio of forces but not by their absolute values. Indeed, the \eq{QHC} for the Hall conductivity contains only the ratio $v_{cm}/v_L$.


\section{Hydrodynamics of superfluids in periodic potentials} \label{CML}

Hydrodynamics of superfluids in periodic potentials  after coarse-graining (averaging over potential periods) reduces to the continuous model with the Lagrangian:\cite{Son13a,EBS}
\begin{equation}
L=- \hbar n \dot \theta 
 - {\hbar^2  \tilde n \over 2m}(\bm \nabla \theta )^2 -E_c(n). 
                   \label{contL} \end{equation}
We consider a 2D problem, where $n$ is the particle number per unit area, and $E_c(n)$ is the energy of a resting liquid which depends only on  $n$.
The Hamiltonian (energy) for this Lagrangian is
\begin{equation}
{\cal H}=  {\partial L\over \partial \dot \theta} \dot \theta -L={\hbar^2  \tilde n \over 2m}(\bm \nabla \theta )^2+E_c(n) .
                   \label{contHh} \end{equation}
Despite similarity of the model to hydrodynamics of the perfect fluid, there is an essential difference. Averaging over the potential period restores translational invariance but not Galilean invariance. The latter is absent since the effective density $\tilde n$, which characterizes stiffness of the phase field,  is different from the true particle density $n$ and can be much smaller than $n$.

According to  Noether's theorem, the gauge invariance provides the conservation law for charge (particle number):
\be
{\partial \over \partial t}{\partial L\over \partial \dot \theta} +\nabla_k\left({\partial L\over \partial \nabla_k \theta} \right)=0. 
    \ee{}
This  is the continuity equation (the first Hamilton equation) for the fluid:
\be
m{\partial n \over \partial  t}=-\bm \nabla \cdot {\bm j} .
    \ee{nt}
The mass current
\be
\bm j=-{m\over \hbar}{\partial L\over \partial \bm\nabla \theta} =\hbar \tilde n\bm \nabla \theta
   \ee{Gcur}
 by the factor $m/q$ differs from the charge current of particles with the charge $q$.  

The second Hamilton equation for the phase $\theta$ canonically conjugate to $n$ (Josephson equation) is
\be
\hbar {\partial \theta \over \partial t}= -{\partial H\over \partial n }=-m\mu=-m\mu_0-{d\tilde n\over dn} {\hbar^2 (\bm \nabla \theta)^2\over 2m }  ,
         \ee{nph}
where $\mu_0 =\partial E_c(n)/m\partial n$ is the chemical potential $\mu$ of the fluid at rest.

Noether's theorem also provides the conservation law
\bem
{\partial g _k \over \partial t} +\nabla_l\tilde\Pi_{kl}=0,
     \eem{NT}
 for the momentum with the density (current)
\be 
\bm g=-{\partial L\over \partial \dot \theta}\bm\nabla  \theta =\hbar n\bm \nabla \theta.
         \ee{}
The conservation law is related to translational invariance of the model.

Here the flux tensor is
\bem 
\tilde  \Pi_{kl}={\partial L\over \partial \nabla_k\theta} \nabla_l\theta-L\delta_{kl}={\hbar^2\over  m  } \tilde n\nabla_k\theta  \nabla_l\theta
 \nonumber \\
+\left[P  +{\hbar^2\over 2m }\left({d \tilde n\over dn} n-\tilde n\right)(\bm \nabla \theta)^2  \right] \delta_{kl},
   \eem{MFl}
The  pressure $P$ is determined by the $T=0$ thermodynamic   relation
\be
P =n{\partial E\over \partial n} - E =n \mu_0-E_c(n) +{\hbar^2\over 2m }\left(n{d\tilde n\over dn}-\tilde n\right)  (\bm \nabla \theta)^2.
    \ee{}
Using Bernoulli's law following from the Josephson equation (\ref{nph}) for the stationary case one can exclude the pressure from \eq{MFl}, and 
 \bem 
\tilde  \Pi_{kl}={\partial L\over \partial \nabla_k\theta} \nabla_l\theta-L\delta_{kl}={\hbar^2\over  m  } \tilde n\nabla_k\theta  \nabla_l\theta
 \nonumber \\
+\left[P_0-\tilde n {\hbar^2(\bm\nabla \theta)^2\over 2 m  } \right]\delta_{kl},
   \eem{MFlb}
where $P_0$ is a constant pressure in the absence of any velocity field, which has no effect on the further analysis.

In the Galilean invariant system the current $\bm g$, which appears  in the Noether conservation law following from the translation invariance, coincides with $\bm j$. But in our case with broken Galilean invariance ($\tilde n \neq n$) the currents $\bm g$ and $\bm j$ differ. The true mass current, which at the same time is true momentum density, is $\bm j$. This is because averaging of the microscopic quantum mechanical current, 
\be
\hat {\bm j}=-{i\hbar \over 2}({\hat \psi}^\dagger \bm \nabla {\hat \psi}  - \bm \nabla{\hat \psi}^\dagger {\hat \psi}) , 
   \ee{qmCur}
yields $\bm j$  but not $\bm g$. Here ${\hat \psi}$ and ${\hat \psi}^\dagger$ are the annihilation and the creation operators normalized to the density.  

One can derive our hydrodynamic model from the Bloch band theory taking into account only quantum states close to the lowest band bottom \cite{Son13a,EBS}. This derivation shows  that 
\be
{\tilde n\over n} ={m\over m^*},
     \ee{}
where $m^*$ is the effective mass of particles at the bottom of the lowest Bloch band.  At the same time, Noether's momentum density $\bm g$ coincides  with the density of particle quasimomentum. Therefore we call the flux tensor $\tilde \Pi_{ij}$ emerging from Noether's theorem [\eq{MFl}]  quasimomentum flux tensor.

For a stationary vortex the phase $\theta_v$ around the vortex axis varies with the gradient
\be
\bm \nabla\theta_v ={[\hat z \times \bm r] \over r^2},
      \ee{Vgrad}
where $\bm r$ the position vector with the origin at the vortex axis. The kinetic energy related with this gradient essentially suppresses the particle density close to the vortex line. Correction $n'$ to the density $n$  is determined from the Josephson equation (\ref{nph}) at $\dot \theta=0$. The distance at which $n'\sim n$ determines the vortex core radius\footnote{In Ref.~\onlinecite{EBS}, Sec.~13.2,  the core radius was defined as a distance at which correction to $\tilde n$ becomes of the order of $\tilde n$, i.e., $n' (\partial \tilde n/\partial n) \sim \tilde n$. In the present analysis, definition of $r_c$ from the condition $n' \sim n$ is more appropriate.} 
\be
r_c  \sim  {\hbar\over  mc_s}\sqrt{d\tilde n\over dn} ,
     \ee{core}
where $c_s= \sqrt{n \partial  \mu_0/\partial n}$ is the sound velocity in a uniform fluid. This estimation makes sense as far as $r_c$ is longer than the period of the periodic potential, to which we apply our model (see below).

Our equations are not invariant with respect to the Galilean transformation. In the coordinate frame moving with the velocity $\bm w$ 
\begin{equation}
H'={\hbar^2  \tilde n \over 2m}(\bm \nabla \theta' )^2+\hbar (\tilde n -n)\bm w \cdot \bm \nabla \theta' +E_c(n) , 
                   \label{HhG} \end{equation}
\be
\bm g'= \hbar n \bm \nabla \theta',~~\bm j' =\hbar \tilde n \bm \nabla \theta' +(\tilde n -n) m \bm w.
       \ee{}
\bem 
\tilde  \Pi'_{kl}={\hbar^2\over  m  } \tilde n\nabla_k\theta ' \nabla_l\theta' +\hbar(\tilde n -n) \nabla \theta'  w_l
 \nonumber \\
+\left[P_0-\tilde n {\hbar^2(\bm\nabla \theta')^2\over 2 m  }-\hbar (\tilde n -n)\bm w \cdot \bm \nabla \theta' \right] \delta_{kl}.
   \eem{MFtr}
The relation between phase gradients in the laboratory and  the moving coordinate frame is $\bm \nabla \theta'=\bm \nabla \theta - m \bm w /\hbar$. 

In a Galilean-invariant superfluid any force on the vortex is determined by  the momentum flux through a cylindric surface surrounding the vortex line \cite{RMP,EBS}. In a superfluid in a periodic potential the true momentum is not conserved since there is  a momentum exchange between the superfluid and the system, which provides the periodic   potential. Instead of it  there is a conservation law for quasimomentum.  Then one may expect that forces on the vortex are determined by the quasimomentum fluxes $\oint  \tilde \Pi_{kl} dS_l$ through a cylindric surface surrounding the vortex. This agrees with the well established result of the Bloch theory for solids that an external force on particles is related with variation of quasimomentum but not true momentum. Let us derive the Magnus and the Lorentz forces on the basis of this assumption.

In the case the Lorentz force the vortex is at rest in the laboratory coordinate frame connected with periodic potential. Then one must use \eq{MFlb} for the quasimomentum flux, where the phase gradient  $\bm\nabla  \theta= \bm \nabla\theta_v  + \bm \nabla\theta_t $ consists of the gradient $\bm \nabla\theta_v $ induced by the vortex line [\eq{Vgrad}] and the gradient $\bm \nabla\theta_t  =\bm j /\hbar  \tilde n$ produced by the transport current. The force arises from the cross terms $\bm \nabla\theta_v  \cdot \bm \nabla\theta_t $ in the quasimomentum flux tensor. Their  integration yields the Lorentz force given by \eq{Lf} The mass current $\bm j =\hbar \tilde n \bm \nabla \theta_t$ is proportional to $\tilde n$, which differs from the particle density $n$.

Calculating the Magnus force proportional to the vortex velocity it is convenient to use the coordinate frame moving  with the velocity $\bm w=\bm v_L$, in which all variables are time-independent. In this frame the quasimomentum flux is determined by \eq{MFtr}. The superfluid is at rest in the laboratory frame, but in the moving frame it moves with the velocity $-\bm v_L$. Then the phase gradient in \eq{MFtr} is $\bm \nabla\theta' =\theta_v -m\bm v_L /\hbar$. Eventually one obtains  the same \eq{Mfg} for the Magnus force like in the Galilean invariant fluid.

Thus the quasimomentum balance yields that the vortex moves with the center-of-mass velocity given by \eq{cmv}. This is a generalization of Helmholtz's theorem for a Galilean invariant perfect fluid, which tells that the vortex moves with the fluid velocity. In general the fluid velocity must be the center-of-mass velocity, but not the superfluid velocity $\bm v_s$. The latter is the average fluid velocity.\cite{EBS} The center-of-mass velocity  coincides with the group velocity $(\hbar/m^*) \bm \nabla \theta$ in the Bloch band theory. For charged fluids the quasimomentum balance confirms the universal quantum Hall conductivity given by \eq{QHC}. 

Validity of derivation of forces on the vortex from the quasimomentum balance is not self-evident.  The vortex in a sense is an alien body immersed into the fluid. 
If one looks for the force between the immersed body and the fluid strictly speaking one cannot rely on the quasimomentum balance because the very concept of the quasimomentum valid only inside the fluid and cannot be used on the other side of the interface between the fluid and the body. For strict justification one should refer to the true momentum balance.

\section{Course-grained Gross--Pitaevskii theory } \label{GPT}

We need to analyze the momentum balance at scales smaller than the core radius $r_c$ given by \eq{core}  at which the hydrodynamic model of Sec.~\ref{CML} is not valid. One must refer to a more general theory, namely the course-grained Gross--Pitaevskii theory, ie., averaged over the potential period.
We shall derive this theory for a weak periodic potential. Although the weak potential weakly affects  forces on the vortex its effect can be calculated exactly.

In the Gross--Pitaevskii theory one can replace the operators ${\hat \psi (\bm r)}$ and ${\hat \psi}^\dagger(\bm r)$ by classical complex-conjugated fields ${ \psi (\bm r)}$ and ${ \psi}^*(\bm r)$ satisfying the nonlinear Schr\"odinger equation
\be
i\hbar \dot \psi= -{\hbar^2\nabla^2 \psi \over 2m} +U(\bm r)\psi +V|\psi|^2\psi,
    \ee{nsh}
where $U(\bm r)$ is an external potential for particles (periodical in our case) and $V$ is the amplitude of particle-particle interaction. The balance equation of non-conserved momentum can be derived from this equation:
  \be
{\partial j_k \over \partial t} + \nabla _l\Pi_{kl} = |\psi |^2 \nabla_k U (\bm r).
      \ee{balGP}
The true momentum flux tensor is 
\bem
\Pi _{kl} = {\hbar^2(\nabla_k \psi^* \nabla_l \psi +\nabla_k \psi \nabla_l \psi^*)\over 2m} 
\nonumber \\
+ \left(- {\hbar^2\nabla^2 |\psi|^2\over 4m} +{V|\psi|^4\over 2}\right)\delta_{kl}.
       \eem{trM}

Let us start from a one-dimensional periodic potential varying only along the axis $x$. In the coordinate frame moving with the velocity $\bm w$ the Gross--Pitaevskii equation is
\be
i\hbar \dot \psi= -{\hbar^2\nabla^2 \psi \over 2m} +U_0 \cos {2\pi (x-w t)\over a}\psi +V|\psi|^2\psi.
    \ee{se1}
If the particle-particle interaction $\propto V$ is weak the general solution of this equation is a superposition of Bloch functions
\be
\psi_B(\bm r,t) =u_l(\bm r, \bm  k)e^{i \bm k\cdot \bm r-iE(\bm k) t/\hbar},
           \ee{BF}
where $u_l(\bm r, \bm  k)$ are periodic functions with a period $a$. We want to derive a continuous description in terms of the envelope function
\be
\Psi(\bm r, t)= \int \Psi(\bm k, E )e^{i \bm k\cdot \bm r-iE(\bm k)t}\,d\bm k dE,
     \ee{}
which slowly varies on the lattice period $a$. The derivation assumes that only states with small $k$ ($k\ll 1/a$) in the lowest Bloch band $l=0$ are important in the superposition of Bloch functions. Using the perturbation theory one can obtain an expression for the microscopic wave function not averaged over the period:
\bem
\psi =\Psi  \left(1-{u\over 2}\cos {2\pi x'\over a}+u{i  m wa \over 2\pi \hbar}\sin {2\pi x'\over a}\right) 
\nonumber \\
 +{ua\over 2\pi}\nabla_x \Psi  \left(  \sin  {2\pi x'\over a} +{2im wa \over \pi \hbar}\cos {2\pi x'\over a} \right) 
 \nonumber \\
 +{u a^2\over 2\pi^2} \left(\nabla_x^2 \Psi 
   + {mV  |\Psi |^2 \over \hbar^2}\Psi \right)\cos {2\pi x'\over a},
        \eem{wfm}
where $x'=x-w t$ and $u=ma^2 U_0 /\pi^2\hbar^2$ is the parameter of the perturbation theory. One can check by substitution that  if the envelope function satisfies the coarse-grained Gross--Pitaevskii equation,
\bem
i \hbar \dot \Psi=-{\hbar^2 \nabla_x^2 \Psi \over 2m}\left(1 - {u^2\over 2}\right)-{\hbar^2 \nabla_y^2 \Psi \over 2m} 
\nonumber \\
+{ \hbar  u^2  \over 2} i w   \nabla_x \Psi +\tilde V|\Psi|^2 \Psi ,
    \eem{}
the microscopic wave function (\ref{wfm}) satisfies the original microscopic Gross--Pitaevskii equation (\ref{se1}) neglecting the gradients of $\Psi $ higher than of second order and higher harmonics of the periodic potential with wave numbers $2\pi l/a$ at integer $l>1$.  Here
\be
\tilde V=V\left(1+{u^2\over 2}\right).
   \ee{}

The Gross--Pitaevskii equation for the envelope function $\Psi$ corresponds to the Lagrangian 
\bem
L_v ={i\hbar \over 2} (\Psi^* \dot \Psi -\Psi \dot \Psi ^*)-{\cal H} ,
         \eem{}
where 
\bem
{\cal H}={\hbar ^2\nabla_x ^2| \Psi|^2\over 2m} \left(1 - {u^2\over 2}\right)+{\hbar ^2\nabla _y ^2| \Psi|^2\over 2m}\nonumber \\
 +{i\hbar u^2 \over 4} w (\Psi^*\nabla_x \Psi -\Psi \nabla_x \Psi ^*)
   +{\tilde V|\Psi|^4\over 2}
       \eem{}
is the Hamiltonian.

From Noether's theorem one obtains the conservation law for the quasimomentum density $\bm g$ 
\be
\bm g=-{\partial L_v\over \partial \dot\Psi} \bm \nabla \Psi +{\partial L_v\over \partial \dot\Psi^*} \bm \nabla  \Psi^*=-{i\hbar\over 2}  \left(\Psi^* \bm\nabla \Psi -\Psi \bm \nabla  \Psi^*\right)
  \ee{}
with components of  the quasimomentum flux tensor 
\bem 
\tilde \Pi_{xx}={\hbar^2\over m}|\nabla_x  \Psi|^2 \left(1 - {u^2\over 2}\right)
\nonumber \\
 +{i\hbar u^2 \over 4} w (\Psi^*\nabla_x \Psi -\Psi \nabla_x \Psi ^*)
 \nonumber \\
-{\hbar^2\nabla_x ^2| \Psi|^2\over 4m} \left(1 - {u^2\over 2}\right)-{\hbar^2\nabla _y ^2| \Psi|^2\over 4m} +{\tilde V|\Psi|^4\over 2}, 
      \eem{}
\bem 
\tilde \Pi_{yy}={\hbar^2\over m}|\nabla_x  \Psi|^2 
 \nonumber \\
-{\hbar^2\nabla_x ^2| \Psi|^2\over 4m} \left(1 - {u^2\over 2}\right)-{\hbar^2\nabla _y ^2| \Psi|^2\over 4m} +{\tilde V|\Psi|^4\over 2}, 
      \eem{}
\bem 
\tilde \Pi_{xy}={\hbar^2\over 2m}(\nabla _x\Psi^*\nabla_y  \Psi+\nabla _x\Psi\nabla_y  \Psi^*),
    \eem{}
\bem 
\tilde \Pi_{yx}={\hbar^2\over 2m}(\nabla _x\Psi^*\nabla_y  \Psi+\nabla _x\Psi\nabla_y  \Psi^*)\left(1 - {i\hbar u^2 \over 4}\right)
\nonumber \\
 +{i\hbar u^2 \over 4} w (\Psi^*\nabla_y \Psi -\Psi \nabla_y \Psi ^*).
     \eem{}
The flux tensor is not symmetric because cylindric symmetry is absent.   

The theory is easily extended to more general harmonic periodic potentials, since in the perturbation theory up to the second order in the potential $U_0$ contributions from separate harmonic potentials are additive. For periodic potentials of square,
\be
U(\bm r)=U_0 \left(\cos {2\pi x\over a}+\cos {2\pi y\over a}\right),
   \ee{u4}
or hexagonal symmetry,  
\bem
U(\bm r)=U_0 \left[\cos {2\pi x\over a}
\right. \nonumber \\ \left.
+\cos {2\pi (\sqrt{3}x +y) \over 2a}+\cos {2\pi (\sqrt{3}x -y) \over 2a}\right],
   \eem{u6}
the coarse-graining  Gross--Pitaevskii equation in the coordinate frame moving the velocity $\bm w$ is 
\be
i \dot \Psi=-{\hbar^2 \bm \nabla ^2 \Psi \over 2m^*}+\left(1-{m\over m^*}\right) i \hbar \bm w \cdot \bm  \nabla  \Psi +\tilde V|\Psi|^2\Psi.
    \ee{GPgr}
This corresponds to the Hamiltonian
\bem
{\cal H}={\hbar ^2| \bm \nabla  \Psi|^2\over 2m^*} 
\nonumber \\
 +\left(1-{m\over m^*}\right){i\hbar  \over 2} \bm w\cdot (\Psi^*\bm \nabla  \Psi -\Psi \bm \nabla  \Psi ^*)
   +{\tilde V|\Psi|^4\over 2}.
       \eem{}
The mass current and the quasimomentum flux are
\be
\bm j = -{i \hbar (\Psi^*\bm \nabla  \Psi-\Psi\bm \nabla  \Psi^*) \over 2m^*}-\left(1-{m\over m^*}\right)m|\Psi|^2 \bm w,
      \ee{Curr}
\bem
\tilde \Pi_{kl}= {\hbar^2(\nabla_k \Psi^* \nabla_l \Psi +\nabla_k \Psi \nabla_l \Psi^*)\over 2m^*} 
\nonumber \\
 +\left(1-{m\over m^*}\right){i\hbar  \over 2}  (\Psi^*\nabla_k  \Psi-\Psi\nabla _k \Psi^*)w_l
\nonumber \\
+ \left(- {\hbar^2\nabla^2 |\Psi|^2\over 4m^*} +{\tilde V|\Psi|^4\over 2}\right)\delta_{kl}.
       \eem{flux}
For square symmetry
\be
{m\over m^*}=1 - {u^2\over 2},~~\tilde V=V\left(1+u^2\right),
    \ee{}
and for hexagonal symmetry    
\be
{m\over m^*}=1 - {3u^2\over 4},~~\tilde V=V\left(1+{3u^2\over 2}\right).
    \ee{}

After Madelung transformation $\Psi=\sqrt{n}e^{i\theta}$ the coarse-grained Gross--Pitaevskii theory reduces to the hydrodynamic theory of Sec.~\ref{CML} with $E_c(n)=\tilde Vn^2/2$  neglecting contributions from density gradients.

Although Noether's theorem does not lead to the conservation law for the true momentum, the balance equation for the true momentum, which can be derived from the Hamilton equations (\ref{nt}) and (\ref{nph}) neglecting gradient terms of more than second order, looks as a conservation law:
\bem
{\partial j _k \over \partial t} +\nabla_l \pi_{kl}=0.
     \eem{jk}
Here the momentum-flux tensor is
\bem 
\pi_{kl}={\hbar^2\over  m  } {d \tilde n\over dn}\tilde n\nabla_k\theta \nabla_l\theta
+\tilde P \delta_{kl},
   \eem{MFg}
and the partial pressure $\tilde P $ is determined by the relation $d\tilde P =\tilde n d\mu$. Earlier  it was suggested to use this momentum flux tensor for calculation of the Magnus force.\cite{Son13a,EBS} The present analysis has not supported this approach, because the flux tensor $\pi_{kl}$ differs from the microscopical momentum flux tensor (\ref{trM}) averaged over the potential period. This is because the force on the right-hand side of the balance equation  (\ref{balGP}) after averaging can also be reduced to a divergence of some flux, which becomes a part of the momentum flux $\pi_{kl}$. Thus derivation of the momentum balance from coarse-grained equation is not satisfactory since this introduces an uncontrolled  constant flux. Further we shall use the momentum flux derived from equations for a microscopical wave function $\psi$.

\section{Motion of a thick cylinder with velocity circulation around it} \label{thick}

In a Galilean invariant fluid the Lorentz and the Magnus forces on the vortex do nor differ from these forces on a cylinder of the radius much larger than the vortex core radius and with velocity circulation around it moving through the perfect fluid \cite{EBS}. The cylinder can be considered as an artificial vortex core of very large radius. 
It would be useful, at least for pedagogical purposes, to consider  motion of of large-radius cylinder with velocity circulation around it though a superfluid put into a periodic potential. A benefit of this problem is that it allows an analytical solution, while the nonlinear Gross--Pitaevskii theory for a real vortex  has no analytical solution even in the absence of the periodic potential. We shall see that neither the hydrodynamic approach of Sec.~\ref{CML},    nor the more general course-grained Gross--Pitaevskii theory  of Sec.~\ref{GPT} are sufficient for this problem,   because one should look for the momentum flux at the very surface of the cylinder where it is necessary to know the microscopic wave function at scales less than the potential period $a$.

\begin{figure}
\includegraphics[width=.4\textwidth]{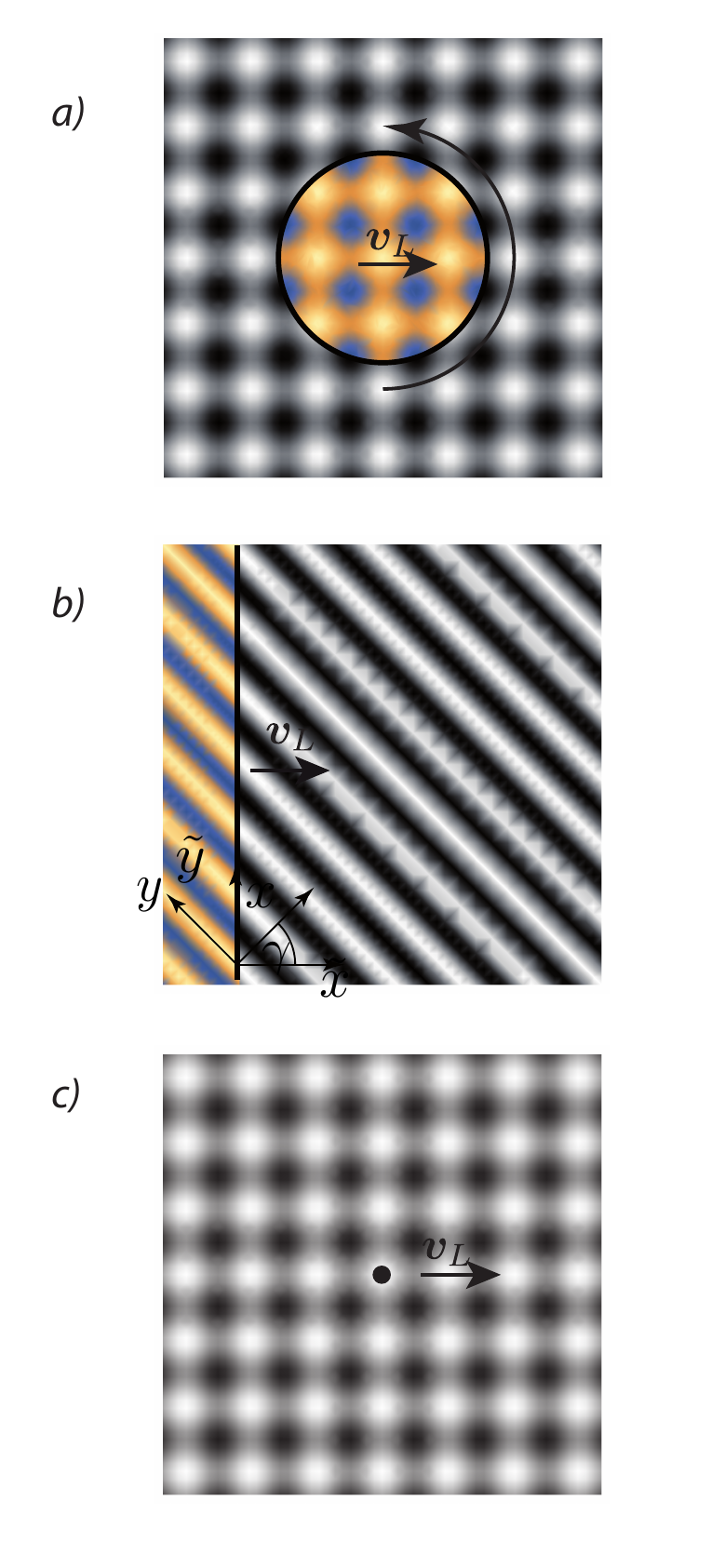}
\caption[]{ Motion of a cylinder with velocity circulation around it through a superfluid in a periodic potential. a) Motion of the cylinder of the radius much larger than the potential period and the vortex core radius. b) Motion of the plain wall through a superfluid in a one-dimensional periodic potential. c) Motion of the cylinder of the radius much smaller than the potential period.
}
\label{f1}
\end{figure}

A cylinder of large radius moves with the velocity $\bm v_L$ through the superfluid  (Fig.~\ref{f1}a) but does not interact with the periodic potential being fully insensitive to it. There is circular superflow around the cylinder with the velocity circulation $h/m$ of a single-quantum vortex.  The cylinder is impenetrable for superfluid, and the microscopic wave function must vanish at its surface. 

Since the cylinder radius  is much larger than the coherence length, at which the wave function  varies from zero at the cylinder surface to its bulk value (healing length), the curvature of the cylindric surface can be ignored.
Therefore it is sufficient to study the case of a plane wall restricting the superfluid and moving with the respect to the periodic potential  (Fig.~\ref{f1}b). In the perturbation theory the contributions from various harmonics in the periodic potential [see Eqs.~(\ref{u4}) and  (\ref{u6})] are additive, and we consider only one harmonics with the periodic potential varying only along the axis $x$ as in \eq{se1}.

The normal to the wall is at the angle $\gamma$ to the axis $x$. After rotation to the new coordinate frame,
\be
\tilde x =x\cos \gamma -y\sin \gamma, ~~\tilde y=x\sin \gamma +y\cos \gamma,
    \ee{} 
the Gross--Pitaevskii equation in the coordinate frame moving with the velocity $v_L$ of the wall,
\bem
i \dot \Psi=-{\hbar^2 \nabla_{\tilde x}^2 \Psi \over 2m}\left(1 - {u^2\over 2}\cos^2\gamma\right)  
\nonumber \\
+{i \hbar  u^2  \over 2} i v_L  \cos\gamma \nabla_{\tilde x} \Psi +\tilde V|\Psi|^2 \Psi ,
    \eem{}
becomes one-dimensional.
The solution of the stationary equation ($\dot \Psi=0$) with vanishing $\Psi$ at $\tilde x=0$ is 
\be
\Psi =\sqrt{n_0}e^{ iu^2 m v_L \tilde x \cos\gamma /2\hbar} \tanh {\tilde x\over \tilde \xi},
      \ee{enSol} 
where $n_0$ is the particle density far from the wall and the effective healing length $\tilde \xi$ is determined by the relation 
\be 
{1\over \tilde \xi^2}= {\tilde V m n_0 \over \hbar^2}\left(1 - {u^2\over 2}\cos^2\gamma\right).
      \ee{}
The phase factor in \eq{enSol}  provides that in the coordinate frame moving with the wall the mass current normal to the wall is absent. However, at the analysis of the healing layer near the wall the velocity $v_L$ of the moving periodic potential yields only inessential quadratic corrections and will be ignored.

The force on the wall from the superfluid is given by the true momentum flux component  $\Pi _{\tilde x \tilde x}$ exactly at the wall, $\tilde x=0$, where the wave function must vanish. Near the wall one can linearize the Gross--Pitaevskii equation (\ref{se1}) neglecting interaction $\propto V$. Then the momentum flux is
\be
\Pi _{\tilde x \tilde x}={\hbar^2|\nabla_{\tilde x} \psi|^2\over m} - {\hbar^2\nabla^2 |\psi|^2\over 4m}.
   \ee{momFpl}
If the force on the cylinder were determined by the quasimomentum flux, the true momentum flux at the cylinder surface would coincide  with the constant quasimomentum flux, which far from the wall is equal to 
\be
\tilde \Pi _{\tilde x \tilde x} ={\tilde V|\Psi|^4\over 2}.
     \ee{qmomFpl}
However, the momentum flux (\ref{momFpl}) differs from the quasimomentum flux (\ref{qmomFpl}) if one uses the microscopic wave function (\ref{wfm}) with the envelope function $\Psi$ given by \eq{enSol}.
This is because a vanishing  at the wall envelope function $\Psi$ does not mean that the microscopic wave function $\psi$ vanishes exactly. Indeed, according to \eq{wfm} close to the wall the wave function $\psi$ is
\be
\psi ={ua\over 2\pi}\cos \gamma \nabla_{\tilde x} \Psi   \sin  {2\pi x'\over a} .
         \ee{wf0}
It is necessary to correct the solution of the linear Gross--Pitaevskii equation (\ref{se1}) for the wave function $\psi$ by adding another solution compensating this error. One can check by substitution that the wave function \begin{widetext}
\bem
\psi' =-{ua\over 2\pi}\cos \gamma \nabla_{\tilde x} e^{-2\pi\tilde x \cos \gamma /a}  \left[\sin   {2\pi \tilde y \sin\gamma\over a} + {u\over 4} \left(\cos 2\gamma \sin {2\pi \tilde x \cos\gamma\over a}  -\sin 2\gamma \cos {2\pi \tilde x \cos\gamma\over a} \right)  \right]
  \eem{}   \end{widetext}
satisfies this equation up to the terms of the second order in $U_0$ and exactly satisfies the boundary condition $\psi'=0$. 
Using the corrected wave function  in the momentum flux (\ref{momFpl})   one obtains that 
the true momentum flux $\Pi _{\tilde x \tilde x}$ is exactly equal  to the quasimomentum flux (\ref{qmomFpl}).
This justifies derivation of forces from the quasimomentum balance. For the sake of simplicity we ignored possible  flow of a superfluid along the wall, but taking it into account does not affect the conclusion. Note  that in order to satisfy the boundary at the wall it is  not sufficient to solve only the coarse-grained Gross--Pitaevskii equation for the envelope function because the  correct wave function close to the wall contains components strongly varying on the scale of the lattice constant.

Thus the cylinder of large radius moves through the superfluid in a periodic potential with the center-of-mass velocity in agreement with Helmholtz's theorem. But for the vortex with small core this conclusion is not true in general, as we shall further see.

\section{Motion of a cylinder with radius less than the potential period } \label{StrInt}

  Let us now consider an opposite limit (still, within the perturbation theory for a weak periodic potential) when the coherence length $\xi = \hbar/\sqrt{Vmn_0}$ is much smaller than the lattice constant $a$.  For satisfying this condition the interaction parameter $V$ must be large enough.
   Here $n_0$ is the average density. We consider a weak periodic potential of square symmetry. The non-linear Schr\"odinger equation (Gross--Pitaevskii equation) in the laboratory frame is
\be
i\hbar \dot \psi= -{\hbar^2\nabla^2 \psi \over 2m} +U_0 \left(\cos {2\pi x\over a}+\cos {2\pi y\over a}\right)\psi 
+V|\psi|^2\psi.
    \ee{Sc}

Since  the vortex core radius $\sim \xi$ is much  smaller than the potential period      the hydrodynamic approach of Sec.~\ref{CML} is sufficient. The density distribution can be found in the Thomas--Fermi approximation, and for the weak periodic potential
\be
n(\bm r) =n_0\left[1 -\tilde u \left(\cos {2\pi x\over a}+\cos {2\pi y\over a}\right)\right],  
   \ee{}
where   $\tilde u=U_0/Vn_0$ is a new perturbation theory parameter.

Suppose that there is a transport superfluid current with the average velocity $\bm v_s =(\hbar/m) \bm \nabla \theta_s$, where $\nabla \theta_s$ is the average phase gradient.  Solving the continuity equation (\ref{nt}) up to the second oder with respect to $\tilde u$,  the phase gradients are
 \begin{widetext}
\bem  
\nabla_x \theta=v_{sx}\left[1+\tilde u\cos  {2\pi x\over a} +{\tilde u^2\over 2}\left(\cos  {4\pi x\over a}+ \cos   {2\pi x\over a}\cos  {2\pi y\over a}\right)\right] 
- v_{sy}{\tilde u^2\over 2}\sin   {2\pi x\over a}\sin  {2\pi y\over a}, 
\nonumber \\
\nabla_y \theta=v_{sy}\left[1+\tilde u\cos   {2\pi y\over a} +{\tilde u^2\over 2}\left(\cos  {4\pi y\over a}+ \cos   {2\pi x\over a}\cos  {2\pi y\over a}\right)\right] 
- v_{sx}{\tilde u^2\over 2}\sin   {2\pi x\over a}\sin  {2\pi y\over a}. 
   \eem{}
The components of the spatially varying mass current are given by 
\bem  
j_x =\hbar n\nabla_x \theta=mn_0v_{sx}\left[1-\tilde u\cos {2\pi y\over a}-{\tilde u^2\over 2}\left(1+ \cos  {2\pi x\over a}\cos  {2\pi y\over a}\right) \right]+ mn_0v_{sy}{\tilde u^2\over 2}\sin  {2\pi x\over a}\sin  {2\pi y\over a},
\nonumber \\
j_y=\hbar n\nabla_y \theta= mn_0v_{sy}\left[1-\tilde u\cos {2\pi x\over a}-{\tilde u^2\over 2}\left(1+\cos  {2\pi x\over a}\cos  {2\pi y\over a}\right)\right] +mn_0 v_{sx}{\tilde u^2\over 2}\sin  {2\pi x\over a}\sin  {2\pi y\over a}.  
   \eem{}
\end{widetext}
The current averaged over the whole plane is 
\bem  
\bm j =mn_0\bm v_{s}\left(1-{\tilde u^2\over 2}\right).  
   \eem{jav}
This points out that the effective mass in our case is
\be
m^*=m\left(1-{\tilde u^2\over 2}\right).
      \ee{} 

In a periodic potential the energy of the vortex (with or without a cylinder in its core)  depends on its position. This produces a pinning force, which pin the vortex to the position, where the energy is minimal. The force on a cylinder with circulation around it produced by a weak gradient of the potential was found in Sec.~1.5 of Ref.~\onlinecite{EBS}: 
\be
\bm F_p= -{\pi \hbar^2 n \over m^2 c_s^2} \ln {a\over r_c} \bm \nabla U(\bm r).
      \ee{forPin}
Here $r_c$ is the radius of an artificial core equal to the cylinder radius but not that given by \eq{core}.
Thus the energy, which pins a cylinder or a vortex to the minimum of the particle density, is 
\be
E_p =- {\pi \hbar^2 n \over m^2 c_s^2} \ln {a\over r_c}  U(\bm r).
     \ee{enPin}
At the  energy   minimum  the external potential $U(\bm r)$ is maximal because the energy is proportional  to the  particle density, which is minimal at the maximum of  $U(\bm r)$.  In the superconductivity theory pinning by a periodic potential in a crystal was  called {\em intrinsic} because it is present even in an ideal crystal, in contrast to extrinsic pinning related with crystal defects.  

Because of pinning  the Lorentz force proportional to the  current is able to move the vortex only if the current exceeds the so-called depinning current.\footnote{Strictly speaking this statement is not accurate because so called quantum vortex slip is possible when vortices jump from one potential well to another via quantum tunneling. But for our problems it is not so important, and anyway, regimes of vortex motion below and above depinning threshold would be different.}
  At currents slightly exceeding the depinning current dependence of the vortex velocity on the current ceases to be linear and vortex motion is rather irregular. However, the concept of linear dependance of  force on current becomes valid again at currents much larger than depinning current. But even in this regime some effect of pinning still remains. Not all trajectories of vortices are equivalent. There are trajectories along which vortices encounter weaker pinning forces. Preferable trajectories go along valleys between ``hills'' of the potential. At vortex motion along the axis $x$ preferable trajectories correspond to discreet values $y=l a$ where $l$ is an integer.  In the superconductivity theory this phenomenon was called guided motion of vortices. Mostly it was investigated not for intrinsic pinning but for extrinsic pinning by defects.\cite{Shkl}

For motion along a preferable trajectory the Lorentz and the Magnus forces are determined by the density $n$ and the  current $\bm j$ averaged not over the whole plane but only over trajectory points:
\be  
\bar n =n_0\left(1 -\tilde u  \right),
~~\bar j=n_0v_s \left(1-\tilde u-{\tilde u^2\over 2}\right) .
       \ee{}
The balance of forces, $\bm F_L+\bm F_M=0$, yields the relation between the vortex and fluid velocities:
\be
v_L\approx v_s \left(1-{\tilde u^2\over 2} -{\tilde u^3\over 2}\right)  \approx {m\over m^*} v_s \left(1 -{\tilde u^3\over 2}\right).
   \ee{}
This differs from Helmholtz's theorem by terms of the third order in $\tilde u$. Although we neglected these terms above more accurate algebra taking into account these terms have not revealed other 3rd order terms in this relation. 
Thus Helmholtz's theorem is not valid, at least in its original formulation. One may still talk at some {\em local} Helmholtz's theorem related to local currents and velocities.

This simple  analysis demonstrates when the approach based on the quasimomentum balance can be wrong. It misses to take into account important features of vortex dynamics, namely those related with intrinsic pinning and guided vortex motion.

\section{Bose--Hubbard model}  

The Hamiltonian of the  Bose--Hubbard model \cite{Ued} for a lattice with distance $a$ between sites is
\be
{\cal H}= -J\sum _{k,l} \hat b_k^\dagger  \hat b_l +{V_N\over 2}\sum_k \hat N_k(\hat N_k-1)- \mu  \sum_i \hat N_k.
    \ee{BH}
Here $\mu$ is the chemical potential, the operators $\hat b_k$ and $\hat b_k^\dagger$ are the operators of annihilation and creation of a boson at the $k$th lattice site,  and $\hat N_k=\hat b_k^\dagger\hat b_k$ is the particle number operator at the same site with integer eigenvalues. The first sum is over neighboring  lattice sites   $k$ and $l$. 

In the superfluid phase with large numbers of particles $N_k$ all operator fields can be replaced by the classical fields in the spirit of the Bogolyubov theory:
\be   
\hat b_k~\to~\sqrt{\langle N \rangle}e^{i\theta_k},~~~\hat b_k^\dagger~\to~\sqrt{\langle N \rangle}e^{-i\theta_k},
        \ee{qaN}
where $\langle N \rangle$ is average number of particles per site and  $\theta_k$ is the phase at the $k$th island. Introducing the pair of canonical variables ``particle number -- phase" the hamiltonian 
(\ref{BH}) becomes a classical hamiltonian
\be
{\cal H}= -E_J\sum _{k,l} e^{i(\theta_l-\theta_k)} +{V_N\over 2}\sum_k  N_k( N_k-1)- \mu  \sum_k N_k,
    \ee{class}
where the Josephson energy is 
\be
E_J=J \langle N \rangle.
    \ee{EJN}

\begin{figure}
\includegraphics[width=.5\textwidth]{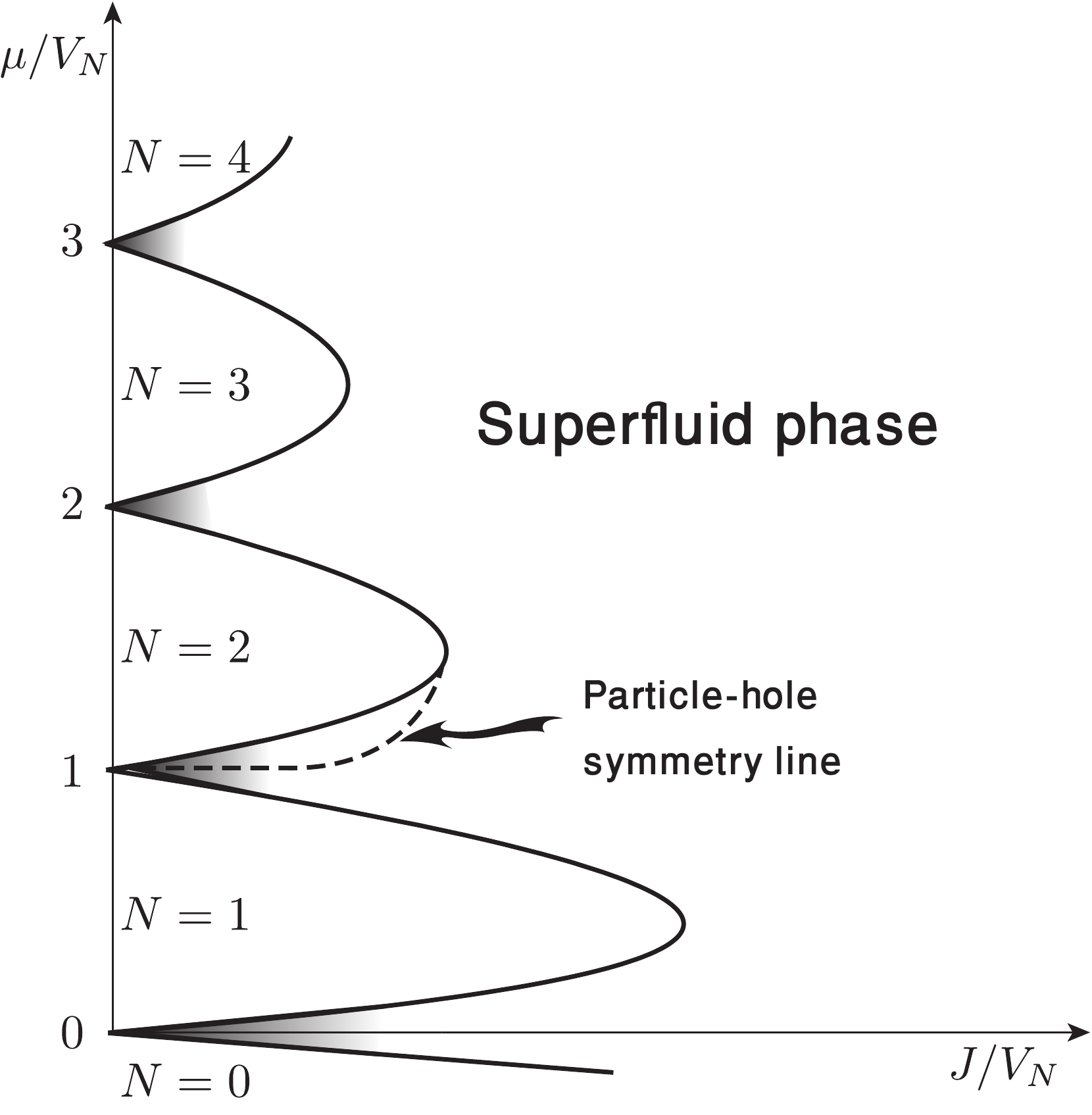}
\caption[]{ The phase diagram of the Bose--Hubbard model.  The Mott insulator phase occupies lobes corresponding to fixed integer numbers $N$ of bosons. The shaded beaks of the superfluid phase between insulator lobes are analyzed in the text. The dash line is the particle-hole symmetry line, which separates the region with the inverse Magnus force from the rest of the superfluid phase. The line is schematic since it was really calculated only in the limit $J\to 0$ where it is horizontal. The particle-hole symmetry line  exists under any lobe but is shown only for the beak between the $N=1$ and  $N=2$ lobes. }
\label{f12-0}
\end{figure}

When the energy $J$ of the intersite hopping decreases, the phase transition from superfluid to Mott insulator must occur.   In the limit $J/V_N \to 0$ eigenstates are  Fock states $|\Psi_N\rangle= |N\rangle$ with fixed number $N$ of particles at any island. At growing $J$ the transition line can be found in the mean-field approximation \cite{Ued}. The phase diagram is  shown in Fig.~\ref{f12-0}. The Mott-insulator phases with fixed numbers $N$ of particles  per site occupy interiors of lobes at small $J/V_N$.

Close to the phase transition at minimal values of $J$, i.e. at beaks of the superfluid phase between  lobes, which are shaded in  Fig.~\ref{f12-0},  the mean-field approximation is simplified by the fact that only two states with $N$ and $N+1$ particles  interplay in the beak between the lobes $N$ and $N+1$. This is because at $\mu=NV_N$ these two states have the same energy, whereas all  other states are separated by a gap on the order of the high energy $V_N$.  Interference of two Fock states leads to broken gauge invariance connected with the transition to the superfluid phase where  the average value of the annihilation operator (and its complex conjugate the creation operator) do not vanish:
\be   
\langle \hat b_k\rangle=\psi_k =|\psi_k| e^{i\theta_k},~~~ \langle \hat b_k^\dagger\rangle=\psi_k^*   =|\psi_k| e^{-i\theta_k}.
        \ee{qa}
Close to the phase transition $|\psi_k|^2 $ is not equal to $\langle N \rangle$ as   \eq{qaN} assumed but can be much smaller:
\be
|\psi_k|^2=(N+1)\left(\frac{1}{ 4 }-N_e^2\right).
   \ee{corP}
Here 
\be
N_e= \langle N \rangle -N-{1\over 2}.
    \ee{Ne}
As in any second-order phase transition, $\psi$ vanishes at the phase transition lines, where $N_e=\pm {1\over 2}$ and the number of particles reaches $N$ at the lower border and $N+1$ at the upper one. Since only a coherent part of the wave function is responsible for Josephson tunneling between islands in  the hopping term one should replace  in the expression (\ref{EJN})  (but not in the interaction term!)    the average particle  number $\langle N \rangle$  by much smaller $|\psi_k|^2$, and the Josephson energy is 
\be
E_J= J(N+1)\left(\frac{1}{ 4 }-N_e^2\right).
 \ee{JEm}

\section{Beyond the Bose--Hubbard model}

Any lattice model does not allow  to analyze the balance of true momentum, since one needs to know a continuous microscopic wave function in the whole space but not only at sites. Therefore we return to a continuous fluid but in the tight-binding limit (strong potential), for which the  Bose--Hubbard model was derived. 

We suggest  a 2D analog of the Kronig--Penney model. There is a square lattice of cylindric deep potential wells (islands), in which the most of particles are trapped, while between the islands particles are described by evanescent tails described by the linear Schr\"odinger equation. A benefit of this model is that there is no forces on the superfluid in the intersite space, and one can rely on the momentum conservation law.

 Let us consider first a single potential well of radius $r_0$. The potential in  the linear Schr\"odinger equation equation is
\be
U(r)=  \left\{ \begin{array}{cc}
-U_0  & \mbox{at}~r<r_0  \\
0 & \mbox{at}~r>r_0   
\end{array}
     \right.
   \ee{}
 The solution of the linear Schr\"odinger equation for the ground state in the island  is 
\be
\psi=  \left\{ \begin{array}{cc}
\sqrt{n_c}J_0(pr)  & \mbox{at}~r<r_0  \\
A {e^{-r/\zeta} \over \sqrt{r}} & \mbox{at}~r>r_0   
\end{array},
     \right.
   \ee{}
where $J_0(pr)$ is the Bessel function, $n_c$ is the particle density at the island center and
\be
\zeta ={\hbar \over \sqrt {2mU_0 -\hbar^2 p^2}}\approx {\hbar \over \sqrt {2mU_0}}
     \ee{}
is the penetration depth into the intersite space.
Continuity of $\psi$ and its derivative  at $r=r_0$ requires that
\be
\sqrt{n_c}J_0(pr_0) = A {e^{-r_0/\zeta} \over \sqrt{r_0}},~~p \sqrt{n_c}J_1 (pr_0) = {A\over \zeta} {e^{-r_0/\zeta} \over \sqrt{r_0}}.
     \ee{}
At $\zeta \ll r_0$ this yields 
\be
p= {2.4\over r_0},~~A =1.71 \sqrt{N_k\over r_0}{\zeta\over r_0},
   \ee{wf}
where the number of particles in the island is
\be
N_k= 2\pi \int\limits_0^{r_0} n_cJ_0(pr)^2  r\,dr=0.533 n_cr_0^2.
   \ee{}

We can also estimate the energy of particle-particle interaction assuming it being too weak for affecting the density distribution in the island:
\bem
E_i= 2\pi \int\limits_0^{r_0} {V|\psi|^4\over 2}r\,dr=0.0737Vn_c^2r_0^2.
   \eem{}

Now let us consider two islands with centers located at the points $x=0$, $y= \pm a/2$ with different phases $\theta_\pm$ of the wave functions. The distance $a$ essentially exceeds all other scales, and overlapping of the wave functions generated by two islands is weak. The wave function outside the islands is
\bem
\psi=A {e^{i\theta_+-r_+/\zeta} \over \sqrt{r_+}}+A {e^{i\theta_--r_-/\zeta} \over \sqrt{r_-}},
   \eem{}
where $r_\pm=\sqrt{x^2 +(y\mp a/2)^2}$. Close to the lattice cell center (small $x$ and $y$):
\be
\psi = {2\sqrt{2}A\over \sqrt{a}}e^{-a/2\zeta-x^2/a\zeta}\left(\cos {\theta\over 2}\cosh{ y\over \zeta} +i\sin {\theta\over 2}\sinh{ y\over \zeta}\right).
     \ee{}
The phase difference $\theta =\theta_+-\theta_-$ leads to the Josephson current between islands:
\bem
I=\hbar \int\limits_{-\infty}^\infty \mbox{Im}(\psi^* \nabla_y \psi)dx={mE_J\over \hbar}\sin\theta,
      \eem{}
where the Josephson energy
\be
E_J = {2\sqrt{2\pi}A^2\hbar^2\over m\sqrt{a\zeta} }e^{-a/\zeta}= {14.66\hbar^2  N_k\zeta\over mr_0^3} \sqrt{\zeta\over a} e^{-a/\zeta} 
      \ee{EJ}
determines the critical current $I_c=mE_J/\hbar$  of the weak Josephson link between two islands. 

Returning back to the square lattice of superfluid islands it is sufficient to take into account only Josephson links between nearest neighbors. Any link is described by our two-islands case, and we receive the Bose--Hubbard model with $E_J$ given by \eq{EJ} and
\be
V_N={2E_i\over N_k^2}={0.519V\over r_0^2}.
     \ee{}

Up to now we addressed the case when the superfluid is very far from the phase transition. Close to the phase transition one must replace in the expression (\ref{EJ}) $N_k$ by $|\psi_k|^2$ given by \eq{corP}. Our Bose--Hubbard model leads to the coarse-grained hydrodynamics of Sec.~\ref{CML} with the parameters
\be
n={N\over a^2},~~\tilde n= {2m E_J \over \hbar^2}={4\sqrt{2\pi}A^2e^{-a/\zeta}\over \sqrt{a\zeta} },~~E_c(n)={V_Nn^2a^2\over 2}. 
    \ee{}
Close to the phase transition
 \be
\tilde n ={2m\over \hbar^2} J (N+1)\left(\frac{1}{ 4 }-N_e^2\right),
     \ee{tn}
and the derivative
 \be
{d\tilde n\over dn}  =a^2{d\tilde n\over d\langle N\rangle}  =-{2m\over \hbar^2} J (N+1)N_ea^2
     \ee{dtn}
vanishes at the line $N_e=0$. This is the particle-hole symmetry line, at which the Magnus force changes its sign.  In Fig.~\ref{f12-0} it is shown by the dashed line.

\section{Intrinsic pinning and guided motion of  vortices}

Effects of intrinsic pinning and guided motion of vortices are definitely stronger in  stronger periodic potentials. The preferable trajectory for a vortex moving in a lattice of islands  along  the axis $x$ goes in the bottom of the valley  between two rows of neighboring islands. Along this trajectory
 the vortex energy has a minimum if the vortex node (the zero of the wave function) is located in a center of the lattice cell with islands at its corners. Such a vortex has the highest four-folded symmetry. 
The maximum of the energy along the trajectory is at the saddle point in the middle of the Josephson link connecting two neighboring islands. 

Accurate calculation of the pinning force at vortex shift is not possible analytically. Therefore starting from this point we present only qualitative calculations based on dimensional estimations.  
The difference between the maximal and the minimal energies along the trajectory is on the order of $\hbar^2\tilde n /m$. Under natural assumption that the dependence on the vortex displacement $d$ close to the cell center is parabolic  the vortex energy there is about
\be
E_p \sim {\hbar^2 \tilde n\over m} {d^2\over a^2}.
      \ee{epin}
Then the qualitative estimation of the pinning force is
\be
F_p \sim {E_p\over d} \sim {\hbar^2 \tilde n\over m} {d\over a^2}.
      \ee{pin}
This estimation defined  the vortex shift as a displacement of the vortex at large distances from its node (we shall call global shift or displacement) and defined the force as derivative of the energy with respect to this displacement. However, the vortex is not a solid rigid object. Any  shift other than multiple  of $a/2$ also deforms it. This leads to ambiguity of the force definition, which will be discussed later in the paper.
 
\begin{figure}
\includegraphics[width=.5\textwidth]{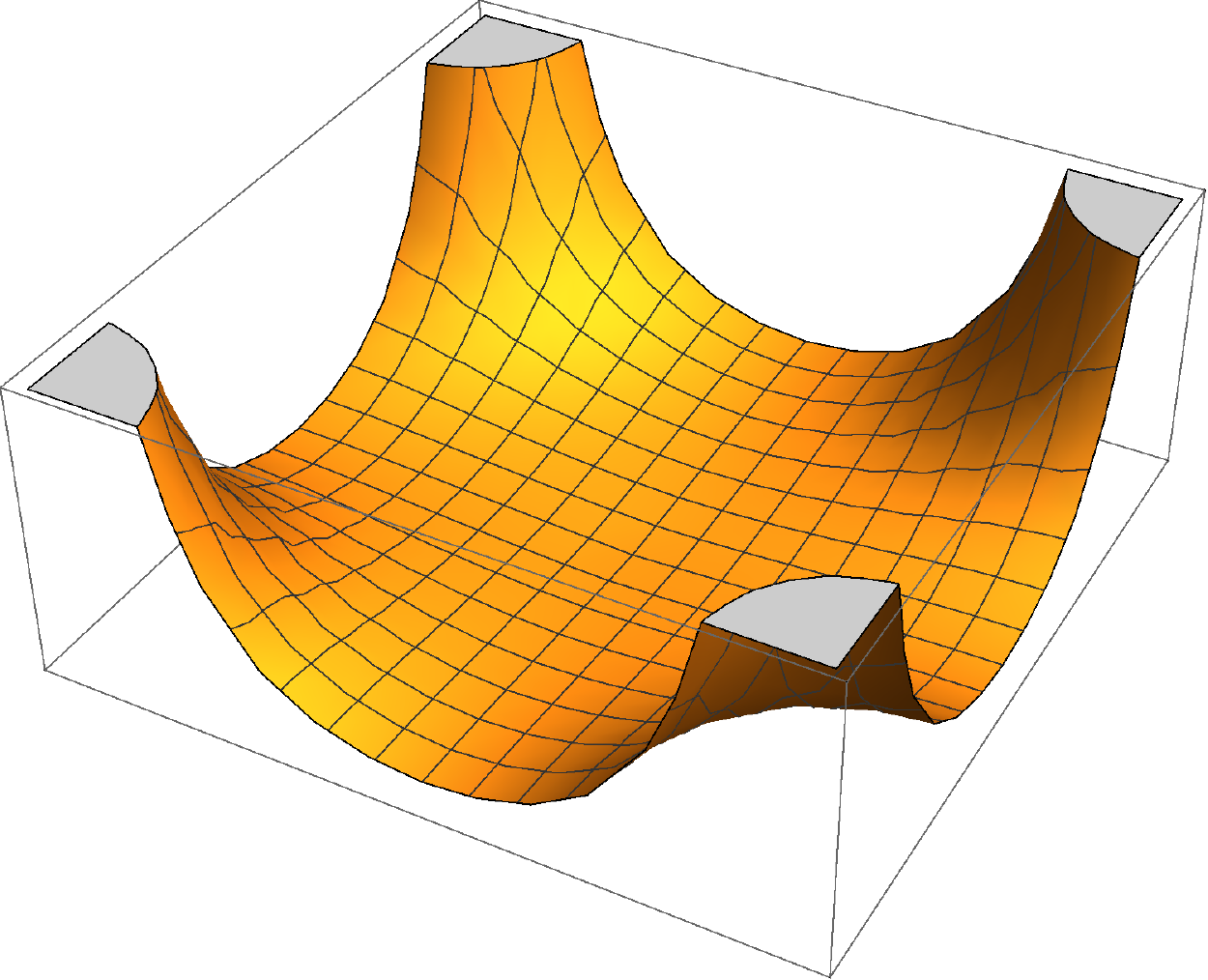}
\caption[]{ Three-dimensional plot for the particle density inside the lattice cell, in which the vortex node is located. }
\label{j3}
\end{figure}

\begin{figure}
\includegraphics[width=.5\textwidth]{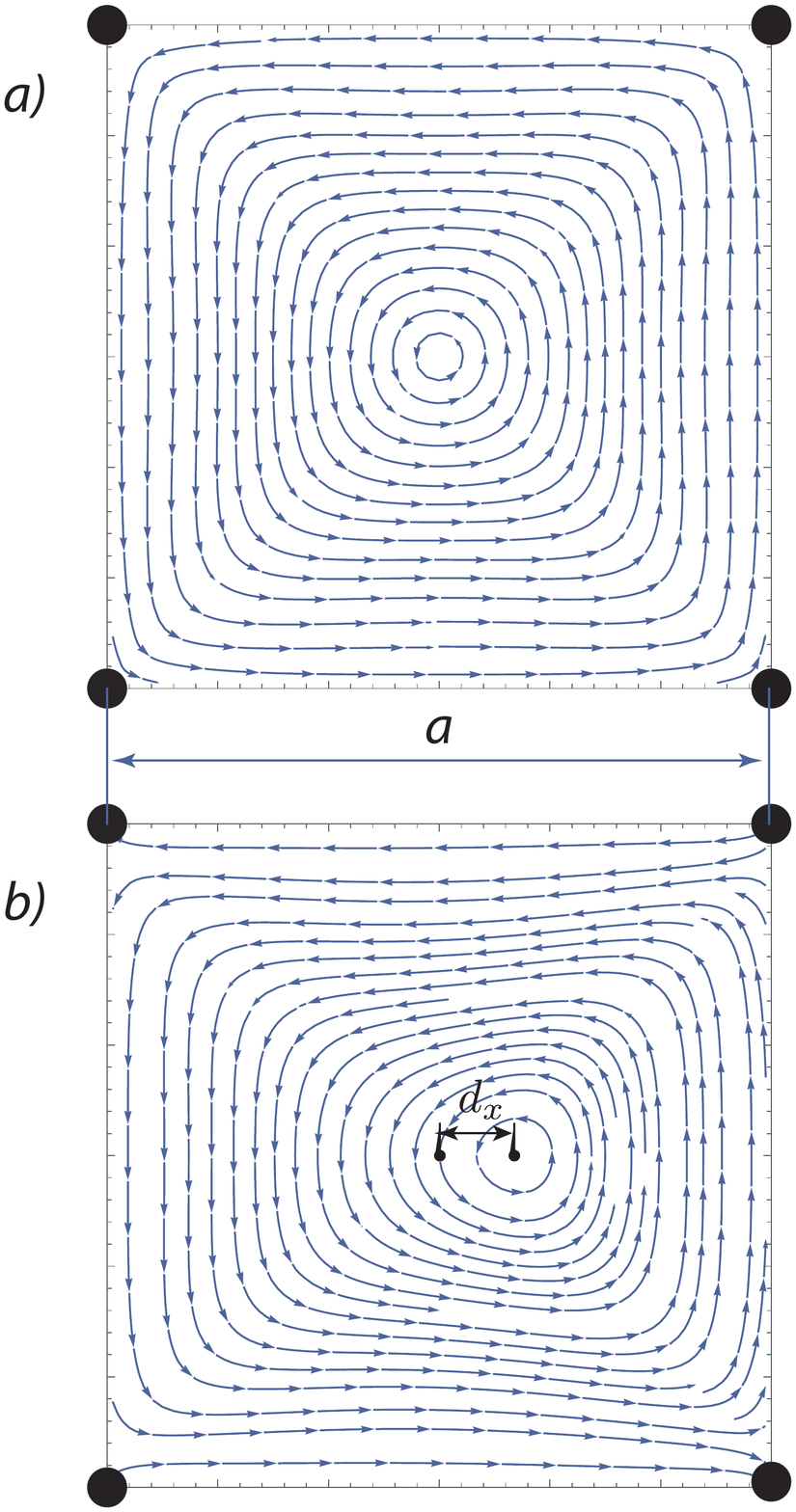}
\caption[]{ Streamlines for the phase gradient $\bm \nabla \theta$ for the wave function $\psi$  inside the lattice cell with the vortex node inside. a) The vortex node is at the center of the lattice cell. b) The vortex node is at the distance $d_x$ from the cell center. }
\label{j2}
\end{figure}

 At the path through the four islands around to the vortex node the phase differences between neighboring islands are $\pi/2$. The wave function in the cell with the vortex node is a linear superposition of wave functions induced by four islands:
\be
\psi= \sum_k  {A_k \over \sqrt{r_k}}e^{i\theta_k-r_k/\zeta},
   \ee{wf4}
where summation is over 4 islands ($k=1,...4$), phases are 
\be
\theta_k =(k-1)\pi/2, 
   \ee{kis}
   and distances from 4 islands are  chosen among 4 values
\bem
r=\sqrt{(x\pm a/2)^2+(y\pm a/2)^2}. 
   \eem{}
Figure \ref{j3} shows the three dimensional plot for the particle density inside the lattice cell, in which the vortex node is located. Figure~\ref{j2}a shows streamlines for the vector field $\bm \nabla \theta$ in the same cell. Here $\theta$ is the phase of the wave function $\psi$.  The vortex node is located in the cell center.
If the particle numbers of of the islands do not differ ($A_k=A$), close to the cell center where $x,y \ll a$ the wave function is
\be 
\psi   ={2\sqrt{2}A\over \sqrt{a}}e^{-a/\sqrt{2}\zeta}\left(-\sinh{x+y\over\sqrt{2}\zeta} +i\sinh{x-y\over\sqrt{2}\zeta}\right).
          \ee{wfc}

We address the case when the superfluid is close to the particle-hole symmetry line when \eq{core} predicts the core radius $r_c$ shorter than the lattice constant $a$. This means that particle numbers in islands  around the vortex node do not differ essentially from those very far from the axis.

Vortex shift $\bm d(d_x,d_y)$  changes the phases of islands around the cell with the vortex node and instead of \eq{kis} they become 
\bem
\theta_1=  -{\pi  (d_x -d_y) \over a},~~ \theta_2={\pi\over 2} -{\pi  (d_x+d_y)  \over  a},
\nonumber \\
 \theta_3=\pi +{\pi  (d_x -d_y)  \over  a},~~ \theta_4={3\pi\over 2} +{\pi  (d_x +d_y)  \over  a}.
     \eem{}
For the vortex  shift along the axis $x$ ($d_y=0$) the wave function transforms from \eq{wfc}    to 
\bem 
\psi ={2\sqrt{2}Ae^{-a/\sqrt{2}\zeta}\over \sqrt{a}}
\left[\sin{\pi d_x \over  a}\left(\cosh{x-y\over\sqrt{2}\zeta} -i\cosh{x+y\over\sqrt{2}\zeta} \right)
\right. \nonumber \\ \left.
-\cos {\pi d_x\over  a}\left( \sinh{x+y\over\sqrt{2}\zeta}  - i \sinh{x-y\over\sqrt{2}\zeta} \right)\right].~~~~~
       \eem{vss}
The node of the vortex (zero of $\psi$) is on the axis $x$ with the coordinate $x_0$ determined by
\be
\tan{\pi d_x\over  a}=\tanh{x_0\over \sqrt{2}\zeta}.
   \ee{node}
Streamlines for the vector field $\bm \nabla\theta$ for the wave function for the vortex node shifted to the distance $d_x$ are shown in Fig.~\ref{j2}b.

 However, at displacements other than multiples of $ a/2$ the state is not stationary, and one need some external force to support it. An external force on the fluid can be applied to a thin  wire of small radius $\rho $ immersed into the fluid at the point with the position vector $\bm r_w(x_w,y_w)$.  We assume the simplest boundary condition that the wave function $\psi$ vanishes at the wire surface. The solution  of the linear Schr\"odinger equation close to the wire is
 \bem
\tilde \psi (\bm r)=\psi  (\bm r)+ \psi(\bm r_w)\left[1- {\ln {|\bm r-\bm r_w|\over \zeta}\over \ln {\rho\over \zeta}}\right]
\nonumber \\
+   \bm \nabla \psi(\bm r_w)\cdot (\bm r-\bm r_w) \left(1-{\rho^2\over |\bm r-\bm r_w|^2}\right),
          \eem{}
where $\psi  (\bm r)$ is the wave function without the wire as before.
The external force on the fluid is determined by an integral of the true momentum flux (\ref{trM}) over the wire surface in which one may neglect the interaction term $\propto V$:
\be 
 \bm F_e={\pi \hbar^2 \over m\ln {\zeta \over \rho} }\bm \nabla |\psi  (\bm r_w)|^2.
        \ee{exF}
This force corresponds to the energy of the wire immersed into the fluid     
\be 
E_e={\pi\hbar^2 \over m\ln {\zeta \over \rho} } |\psi  (\bm r_w)|^2.
        \ee{ewire}
The connection of the global vortex displacement $\bm d$ with  the position vector $\bm r_w$ where the external force is applied to the fluid is derived from the condition that at fixed  $\bm r_w$ the displacement $\bm d$ minimizes the total energy including the large distance  energy (\ref{epin}) of the displaced vortex and the energy (\ref{ewire}) of the wire. This yields the relation
\be
d_x \approx {\sqrt{a\zeta}\over   \ln {\zeta \over \rho}  }e^{-a(\sqrt{2}-1)/\xi}\sinh{x_w\sqrt{2}\over \xi} .
   \ee{wirC}
Using this relation in the expression (\ref{exF}) one obtains the $x$ component of the pinning force:
\be
F_p \sim  {\hbar^2 \tilde n\over m} {d\over a\zeta},
      \ee{pinR}
which by the large factor $a/\zeta$ exceeds the estimation of the pinning force in \eq{pin}. 

Note that the coordinate $x_0$ of the vortex node [\eq{node}] and of the coordinate $x_w$ of the wire [\eq{wirC}] do not coincide, and there is no velocity circulation around the wire. This means that attraction of a vortex to  a defect (thin wire) is not so strong as in the  case of a thick cylinder (Secs.~\ref{thick}   and  \ref{StrInt}), but still is able to push the vortex from cell to cell. At the same time, both $x_0$ and $x_w$  are essentially smaller than the global vortex displacement $d_x$ because of small $\zeta/a$. The disagreement between two estimations (\ref{pin}) and (\ref{pinR}) for the pinning force is connected with the question what displacement is chosen as a measure of the vortex shift. It looks that the global displacement $\bm d$ is a more appropriate choice. The coordinates $x_0$ and $x_w$ describe exponentially small tails far from lattice sites. They, as well as the scale $\zeta$, are hardly relevant  at the macroscopical level. In other words, ``tail does not wag dog''.
Fortunately ambiguity in definition of forces on the vortex does not lead to ambiguous physical conclusions    since in  the most of cases absolute values of the Lorentz and the Magnus forces are not so important. Important is their ratio, or the ratio $v_{cm}/v_L$  of the center-of-mass velocity to the superfluid velocity, which, in particular,  determines the Hall conductivity (\ref{QHC}). Another illustration of this is the case of vortex precession considered in the end of the next section. 

\section{Lorentz and Magnus force in the tight-binding limit} \label{LMF}

As already discussed in Sec.~\ref{StrInt} the Lorentz force is able to make the vortex to move only if the current exceeds the depinning current, and a linear relation between force and current is valid only at currents much larger than the depinning current.  In the lattice models of superfluid, which correspond to tight-binding limit in the Bloch band theory, the problem is even more formidable since depinning velocity and critical velocity are of the same order, and the window where the concept of linear Lorentz force is accurate strictly speaking  is absent. The same problem complicates determination of the  Magnus  force since in a periodical potential the vortex does not move with constant velocity.

In order to overcome this hurdle we redefine usual procedure of determination of the Lorentz and the Magnus force, which allows  to determine them even in the linear theory at very low velocities. We assume that an external force is applied to a cylinder immersed into the fluid as discussed in the previous section. The force is programmed so that it provides a steady motion of the vortex along the prescribed trajectory with constant velocity. A detailed time dependence  of the force and possible effects of quantum tunneling on this dependence are not essential as far as we look for only forces transverse to the trajectory of the vortex.  The low superfluid velocity is not able to make the vortex to move, but it  shifts vortex position. As a result a pinning force emerges proportional to the vortex shift, which must compensate the Lorentz force produced by the superflow. Calculating this pinning force we also obtain the magnitude of the Lorentz force. The same procedure can be used for determination of the Magnus force proportional to the vortex velocity.

The superflow with the velocity $v_s$ parallel to the axis $x$ adds to the island phases $\theta_k$  the same quantity as the displacement $d_y=mv_s a^2 /\hbar$. 
This means that the superflow produces the same force on the cylinder immersed into the fluid as the force (\ref{pin}) for the displacement $d=d_y$. Thus  the magnitude of the Lorentz force is
\be
F_L \sim \hbar \tilde nv_s .
      \ee{}
 This agrees with the Lorentz force   obtained from the quasimomentum balance. Estimating the pinning force by   \eq{pinR} for the same displacement $d_y$ the Lorentz force is by the factor $a/\zeta$ larger.

Let us turn to the Magnus force. If the vortex motion is not accompanied by variation of particle number in islands no transverse force appears. This is a result of particle-hole symmetry.  Indeed, in the coordinate frame moving with the vortex the phases on the islands slightly oscillate but have no constant shift affecting the wave function in the intersite space. 

One can estimate the effect  of particle number   variation in islands with help of the Josephson equation for the Bose--Hubbard model with the hamiltonian (\ref{class}):
\be
\hbar \dot \theta_k ={\partial {\cal H}\over \partial N_k}
= -{\partial E_J\over \partial N_k}\sum _j e^{i(\theta_l-\theta_k)} +V_N N_k- \mu ,
    \ee{}
Close to the particle-hole symmetry line the Josephson-energy term is small,  and at  low velocity $v_L$ the particle number variation at  islands around the vortex node is:
\be
\delta N_k={\hbar \dot \theta_k \over V_N} =  -{\hbar v_L\nabla _x \theta \over V_N} \sim \mp {\hbar v_L \over aV_N},
       \ee{} 
where the upper and the lower signs correspond to the islands above and below the axis $x$ respectively.
Variation of the particle number $N_k$ leads to variation of the parameters $A_k$  in the wave function (\ref{wf4})
\be
A_k \sim A \left( 1+ {\partial E_J \over \partial N_k}{\delta N_k\over E_J}\right ).
     \ee{}
The corrections to the parameter $A$ result in the same corrections of the wave function $\psi$  as those produced by the vortex shift  
\be
d_y \sim a{\partial E_J \over \partial N_k}{\delta N_k\over E_J} \sim {\partial E_J \over \partial N_k}{\hbar v_L\over E_J V_N}\sim {\partial \tilde n  \over \partial n}{\hbar v_L\over \tilde n a^2 V_N} 
    \ee{dFL}
along the axis $y$  [see \eq{vss}]. Equating the  Magnus force to the pinning force (\ref{pin}) one obtains 
\be
F_M  \sim {\partial \tilde n  \over \partial n}{\hbar ^3v_L\over m  a^4 V_N}.
      \ee{FMl}
But estimating the pinning force by the force (\ref{pinR})  on the cylinder immersed into the fluid one obtains by the  factor $a/\zeta$ larger Magnus force.

A detailed analysis of the case far from the particle-hole symmetry line is more  complicated. We restrict ourselves with some extrapolation arguments helping to guess what happens in that case. Our assumption that the particle density in islands close to the vortex line are only weakly suppressed by circular superflows around the vortex becomes invalid when the core radius $r_c$ given by \eq{core} reaches the value of the order of the lattice constant $a$. This correspond to the condition
\be
V_N \sim{\hbar^2\over  m  a^4 n}{d\tilde n\over dn}  .
     \ee{}
Substituting this estimation into \eq{FMl} one obtains the Magnus force
\be
F_M  \sim \hbar n  v_L,
      \ee{}
which qualitatively agrees with that following from the quasimomentum balance. One may expect that this conclusion would be still valid further away from the particle-hole symmetry line.


In conclusion of this section let us consider the phenomenon of vortex precession in the potential well produced by intrinsic pinning.  \Eq{dFL}  is in fact one of the two dynamical equations describing this precession keeping in mind that $\bm v_L =\dot{\bm d}$.  Another equation connects $d_x$ with the $y$ component of the vortex velocity $\bm v_L$.
Using the relations (\ref{tn}) and (\ref{dtn}) valid close to the phase transition the precession frequency is
\be
\omega ={\partial n  \over \partial \tilde  n}{\tilde n a^2 V_N \over \hbar } \sim {V_N \over \hbar N_e}.
     \ee{}
The quantity $N_e$ is given by \eq{Ne}. In the past precession of the vortex in a potential well formed by the trap confining the BEC cloud of cold atoms was discussed theoretically (see Sec.~4.5 of Ref.~\onlinecite{EBS}) and was detected experimentally.\cite{VorPr}
The precession frequency in the intrinsic pinning well can be much higher than the frequency in the trap since the former diverges at the particle-hole symmetry line where $N_e=0$. Observation of the vortex precession in the cold atom BEC in the presence of an optical lattice could be a method to measure the Magnus force experimentally.


\section{Conclusions and discussion}

The paper presented derivation of the Magnus and the Lorentz force on the vortex in a superfluid put in a periodic potential. We considered a weak periodical potential, as well as a strong periodical potential in the tight-binding limit, in which continuous superfluids are frequently described by  lattice models. The latest example is the Bose--Hubbard model describing BEC of cold atoms in optical lattices. 

The starting point of the analysis was the principle argued in Refs.~\onlinecite{Son13a,EBS}: Only the analysis of the balance of the true momentum of the superfluid is a reliable basis for derivation of forces on vortices.  The balance of the quasimomentum as it was defined in the Bloch band theory in solid state physics does not guarantee a correct result in general.  Quasimomentum balance predicts that the vortex moves with the center-of-mass  velocity of the superfluid  in accordance with Helmholtz's theorem. In charged superfluids this yields the universal value of the Hall conductivity [\eq{sHu}], which depends only on particle density, magnetic field, and world constants. Helmholtz's theorem is correct for normal fluids, for an electron gas in a crystal, for example. In superfluids the  theorem sometimes also holds but   not in general  because the superfluid state is not uniform, which results in pinning and guided motion of vortices. 

The analysis demonstrated that the transverse force on the vortex cannot be derived from the coarse-grained description of superfluid, which deals only with envelope wave functions averaged over scales larger than the potential period just because within this description one cannot take into account pinning and guided motion. Neither the transverse force can be derived from lattice models for superfluid, the Bose--Hubbard model in particular,  because they do not contain any information on the wave function in the space between lattice sites. In order to overcome this obstacle  the paper went beyond the Bose--Hubbard model
suggesting some 2D analog of the Kronig--Penney model, from which  the Bose--Hubbard model can be directly derived. 

The theory was applied for a superfluid close to the superfluid--Mott insulator transition. In this part of the phase diagram Lindner {\em et al.}\cite{Auer} and Huber and Lindner\cite{Lind} discovered a line at which the Magnus force changes its sign.  The line was called in the present paper the particle-hole symmetry line. They derived the force from topological arguments connecting the Magnus force with Chern numbers.
 They predicted a sharp transition with a jump in the strength of the Magnus force at  the particle-hole symmetry line. The present analysis has shown that the behavior of the Magnus force at crossing the line is purely analytical  without any jump. This qualitatively agrees with conclusions of Refs.~\onlinecite{Son13a,EBS}, although quantitative expressions for the Magnus force were not correct because an improper expression for the true momentum flux was used (see the end of Sec.~\ref{GPT}).

Instead of dealing with the momentum balance the authors of Refs.~\onlinecite {Auer,Lind,Berg} justified their approach by referring to the Kubo formula for the Hall conductivity derived, in particular, by Avron and Seiler\cite{Avron}. But Avron and Seiler emphasized  that the derivation was valid only if the excitation spectrum has a gap. Oshikawa\cite{Osh} proved the theorem that in a periodic lattice  the gap is  possible only if the particle number per unit cell of the ground state is an integer.  This definitely does not takes place in superfluid states.  The Kubo formula describes a linear response of the current to the electric field, whereas 
 the electric field vanishes  at currents smaller than the depinning current when vortices are pinned.\footnote{A very weak electric field  for currents much less than the depinning current would be possible taking into account quantum tunneling of vortices from one lattice cell to another. But in this case some exponential factors must be present in the final expression. Since they are absent this scenario is also cannot be described by the Kubo formula used in Refs.~\onlinecite {Auer,Lind,Berg}.} Thus the Kubo formula totally ignores effects of intrinsic pinning and guided vortex motion, which played a crucial role in our analysis. 
This alone makes using this formula  in the superfluid state at least questionable. The present paper provides new  evidence that topological arguments not accompanied by the analysis of the momentum balance can be only  guesses for transverse forces justified sometimes in simple cases but not in general.\cite{EBS}  


A possible method of experimental detection of the Magnus force in BEC of cold atoms is observation of precession of a vortex pinned by an optical lattice (Sec.~\ref{LMF}). In a 3D BEC cloud the precession frequency is a gap in the spectrum of Kelvin waves along the vortex line.



%

\end{document}